\documentclass[conference]{IEEEtran}
\usepackage{comment}
\usepackage{amsmath,amssymb,amsfonts}
\usepackage{algorithmic}
\usepackage{cite}
\usepackage{url}
\usepackage{graphicx}
\usepackage{textcomp}
\usepackage{xcolor}
\def\BibTeX{{\rm B\kern-.05em{\sc i\kern-.025em b}\kern-.08em
    T\kern-.1667em\lower.7ex\hbox{E}\kern-.125emX}}
\usepackage{soul} \newcommand{\hll}{}

\newcommand{\replicationPackage}{\url{https://doi.org/10.5281/zenodo.3671463}}

\usepackage{booktabs}
\usepackage{multirow}
\usepackage{multicol}

\usepackage{enumitem}

\usepackage{pgfplots}
\usepackage{pgfplotstable}
\pgfplotsset{compat=1.14}

\newcommand{\q}[1]{\emph{``#1''}}
\newcommand{\qp}[2]{\emph{``#1'' (#2)}}

\usepackage[disable]{todonotes}
\usepackage{todonotes}
\presetkeys    {todonotes}    {inline}{}
    
\usepackage{mfirstuc}
\newcommand{\nfr}[1]{\textsc{#1}}

\usepackage[T1]{fontenc}

\usepackage[many]{tcolorbox}
\newtcolorbox{mybox}[1][]{
  breakable,
  title=#1,
  colback=white,
  colbacktitle=white,
  coltitle=black,
  fonttitle=\bfseries,
  bottomrule=0pt,
  toprule=0pt,
  leftrule=3pt,
  rightrule=3pt,
  titlerule=0pt,
  arc=0pt,
  outer arc=0pt,
  colframe=black,
}

\begin{document}

\title{The Lack of Shared Understanding of Non-Functional Requirements in Continuous Software Engineering: Accidental or Essential?}

\author{\IEEEauthorblockN{
Colin Werner,
Ze Shi Li, 
Neil Ernst, 
Daniela Damian}
\IEEEauthorblockA{
\emph{Department of Computer Science}\\
\emph{University of Victoria, Victoria, Canada}\\
\emph{\{colinwerner, lize, nernst, danielad\}@uvic.ca}}}

\maketitle

\thispagestyle{plain}
\pagestyle{plain}

\begin{abstract}
Building shared understanding of requirements is key to ensuring downstream software activities are efficient and effective.
However, in continuous software engineering (CSE) some lack of shared understanding is an expected, and essential, part of a rapid feedback learning cycle.
At the same time, there is a key trade-off with avoidable costs, such as rework, that come from accidental gaps in shared understanding.
This trade-off is even more challenging for non-functional requirements (NFRs), which have significant implications for product success.
Comprehending and managing NFRs is especially difficult in small, agile organizations.
How such organizations manage shared understanding of NFRs in CSE is understudied.
We conducted a case study of three small organizations scaling up CSE to further understand and identify factors that contribute to lack of shared understanding of NFRs, and its relationship to rework.
Our in-depth analysis identified 41 NFR-related software tasks as rework due to a lack of shared understanding of NFRs.
Of these 41 tasks 78\% were due to avoidable (accidental) lack of shared understanding of NFRs.
Using a mixed-methods approach we identify factors that contribute to lack of shared understanding of NFRs, such as the lack of domain knowledge, rapid pace of change, and cross-organizational communication problems.
We also identify recommended strategies to mitigate lack of shared understanding through more effective management of requirements knowledge in such organizations.
We conclude by discussing the complex relationship between shared understanding of requirements, rework and, CSE.
\end{abstract}

\begin{IEEEkeywords}
shared understanding, 
non-functional requirements, 
continuous software engineering, 
rework
\end{IEEEkeywords}

\section{Introduction}
\label{sec:intro}

Shared understanding is a critical success factor in achieving high quality software that meets stakeholders' needs \cite{bittner_why_2013}.
It also fosters a more effective, motivated, and collaborative team, more efficient use of resources, and less conflict \cite{darch_shared_2009}.
However, empirical research on shared understanding in software engineering projects is scarce \cite{glinz_shared_2015}.

Another critical success factor in software is non-functional requirements (NFR) \cite{wiegers2013software, glinz_non-functional_2007}, \hll{also known as architecturally significant requirements \cite{chen2012characterizing}.}
NFRs, such as \nfr{privacy}, have the potential to derail a software product, for example, if the system violates elements of Europe's General Data Protection Regulation (GDPR).
However, in continuous software engineering (CSE), similar to agile software development, functional requirements (FR) are the primary focus, while NFRs are neglected \cite{ramesh_agile_2010}. 
CSE, especially in small, agile organizations, focuses on automated, rapid release of working software, and often leads to short-term FR prioritization \cite{gralha_evolution_2018}.

One problem with neglecting requirements is requirements-related \textbf{rework}: the extra work needed to fix problems in software due to poorly understood requirements. 
Rework is extremely costly \cite{wagner_literature_2006} and accounts for 40-50\% of the effort on software projects \cite{boehm2005software}.
Some rework may be due to an avoidable lack of shared understanding, which we define as \textbf{accidental} lack of shared understanding. 
In CSE, however, some amount of a lack of shared understanding is \textbf{essential}: inherent in dealing with the essential complexity of software \cite{brooks1987no}, it captures the unknown unknowns \cite{sutcliffe-unknowns} and represents desirable learning and feedback \cite{ries_lean_2011}. 

In this paper we describe empirical evidence from a multi-case, mixed-method study of shared understanding of NFRs in three small, agile organizations employing CSE.
To conceptualize lack of shared understanding of NFRs we traced it to rework development tasks.
Seeking to shed light on the complex relationship between shared understanding, rework, and CSE, our study examined forty-one NFR-related development tasks identified as rework and was driven by the following research questions:

\indent \textbf{RQ1:} What contributes to lack of shared understanding \indent \indent \indent \indent of NFRs? \\
\indent \textbf{RQ2:} Which NFRs are most associated with a lack of \indent \indent \indent \indent shared understanding? \\
\indent \textbf{RQ3:} What amount of a lack of shared understanding of \indent \indent \indent \indent NFRs is \textbf{accidental} versus \textbf{essential}?

In our close, iterative interactions with our industrial partners, we learned what factors contribute to a lack of shared understanding of NFRs, a large proportion of rework was due to an \textbf{accidental} lack of shared understanding, and strategies our collaborators recommended to avoid it. 
Our work brings two significant contributions:
\begin{enumerate}
    \item We add to the scarce, yet much needed, empirical evidence on shared understanding of NFRs in software engineering, including contributing factors and practices towards avoiding it.
    \item We bring awareness, based on empirical findings, to challenges that CSE brings to shared understanding of NFRs.
\end{enumerate}

Finally, our study provides important implications for both research and practice. 
For research, there is a need to address new challenges brought forth by CSE in RE  by revisiting supporting roles, methods, and tools for the identification, documentation, and communication of requirements. For practitioners, we outline how to document and communicate NFRs to avoid a lack of understanding in CSE.

\section{Background and Related Work}
\label{sec:background}

\subsection{Shared Understanding}
Creating a shared understanding is an important aspect and challenge to consider when attempting to develop high-quality software \cite{hoffmann2013emergence}, especially in RE \cite{corvera2013emergence}.
Yet there has been a lack of ``systematic treatment or classification of the different forms of shared understanding, neither in general nor in a software engineering context'' \cite{glinz_shared_2015}.
Glinz and Fricker \cite{glinz_shared_2015} developed a model describing four quadrants of shared understanding consisting of two forms, explicit and implicit shared understanding, and two categories, true and false understanding. 

The most important goal of RE is to create a shared understanding between the development team and stakeholders \cite{fricker2015requirements}; moreover, the practice of creating and maintaining a shared understanding is not well established \cite{schon_agile_2017}.
Previous work has shown that a misunderstanding of requirements at the beginning of development resulted in substantial rework \cite{bjarnason2011case}.
Unfortunately, shared understanding is \hll{\emph{often}} passive, informal, and unstructured \cite{glinz_shared_2015}; the result of which negatively affects software quality and necessitates rework \cite{damiantse2006}, which is further compounded with the cross-cutting nature of NFRs.
A recent survey \cite{capgemini} stipulates creating a shared understanding of customer expectations is the top challenge in agile adoption, yet understudied \cite{glinz_shared_2015}.
However, shared understanding of NFRs is even harder to study, due to the cross-cutting nature of NFRs, and has not been studied in the rapidly changing environments of agile or CSE, yet remains a critical success factor to software and merits further attention.

\subsection{Non-Functional Requirements}
Despite their importance to success \cite{glinz_shared_2015}, NFRs are ambiguous, not well documented, shared in an informal matter, and not well understood \cite{chung2012non, ameller_how_2012}.
The wide-ranging and extensive NaPiRE study \cite{Fernandez2018} found that ``unclear / unmeasurable'' NFRs were one of the top problems respondents had in their small organizations \cite{Wagner:2017aa}. What some organizations think are NFRs are actually FRs \cite{eckhardt_are_2016}.
NFRs are also inherently difficult to verify or validate \cite{borg_bad_2003, jiang_co-evolution_2015}.
NFRs are cross-cutting across multiple disciplines, especially in a short, fast paced software development iterations \cite{bellomo_evolutionary_2014}. 
Finally, the inability to decompose an NFR into measureable and valuable artifacts that can be completed within a release cadence \cite{bellomo_evolutionary_2014} further increases the difficulty in managing an NFR in CSE.
This difficulty impedes the forming of a shared understanding of that NFR and can result in costly rework.
A number of empirical studies have been performed on the challenges of NFRs in practice \cite{berntsson_svensson_quality_2009, alsaqaf_understanding_2018, behutiye_non-functional_2017}; however, these studies do not focus on identifying lack of shared understanding of NFRs in CSE.

\subsection{Rework}
Various definitions and classifications of rework have been proposed, such as Swanson's three classification types (corrective, adaptive, and perfective) \cite{swanson_dimensions_1976} or Basili's classification based on the source of rework \cite{basili_characterizing_1997}.
Regardless of the classification and the large costs associated with rework \cite{basili_characterizing_1997}, there is consensus that some rework could \emph{not} have been prevented \cite{boehm2005software} and is acceptable.
In fact, a certain amount of rework is desirable \cite{fairley_iterative_2005}, as no rework could indicate developers are not performing their jobs with due diligence, popularized in the ideas captured by Ries \cite{riesStartup}. Technical debt due to NFRs increases the number of quality issues, which can worsen and compound rework upon rework in later development phases \cite{martini_danger_2015}.
False implicit shared understanding \cite{glinz_shared_2015} of NFRs can lead to substantial technical debt and ultimately force an organization to perform major rework \cite{behutiye_management_2019}.

\subsection{Continuous Software Engineering}
CSE is a paradigm that emphasizes rapid and automated releases of working software \cite{fitzgerald2017continuous}.
\hll{This suggests that RE, including the treatment of NFRs, are more challenging in CSE, whereby many common best practices may not be followed and requirements may be informally captured ``just-in-time'' \cite{ernst_case_2012}.}

A number of studies \cite{jaatun2018, cukier2013, shahin_empirical_2019, feitelson_development_2013, li_reliability_2011, bellomo_toward_2014} have explicitly studied the management of NFRs in a CSE context; however, none of them focus on the associated lack of shared understanding. 
Feitelson describes how Facebook uses an open source tool, Perflab, to provide metrics which Facebook monitors to evaluate the \nfr{performance} of their system \cite{feitelson_development_2013}. 
While metrics are a key strategy to operationalize and create a shared understanding of an NFR in CSE, Perflab is self-proclaimed as ``not supported'' and ``not turn-key''.

For software \nfr{security}, Jaatun argues that proper attention to incident management, including involding and educating developers, can help alleviate \nfr{security} issues in CSE \cite{jaatun2018}; although this is strictly for \nfr{security} and is primarily focused on the Building Security In Maturity Model \cite{mcgraw2009building}.

Other studies \cite{cukier2013, bellomo_toward_2014, shahin_empirical_2019} examine how to architect a system to be used in CSE, primarily focusing on deploying as a serverless cloud-based platform.
These studies focus on NFRs, such as \nfr{scalability} and \nfr{deployability}, that have significant influence on a system's architecture, including positives, negatives, and trade-offs.
While these studies show how to architect a system in CSE, they do not look at how to create a true shared understanding of explicit NFRs in CSE.
Furthermore, no empirical evidence exists that measures the effect of shared understanding of NFRs \cite{glinz_shared_2015}, let alone in CSE.
Thus, our goal is to fill this gap in research by explicitly focusing on the state of the practice of shared understanding of NFRs in CSE.

\section{Research Methodology}
\label{sec:method}
We conducted a multi-case study \cite{yin_case_2002} using a mixed-methods approach \cite{seaman_qualitative_1999} in collaboration with three independent organizations using qualitative and immersive techniques \cite{potts_research}.
To study lack of shared understanding of NFRs in relation to rework, we investigated how the organizations handled NFRs and specifically traced lack of shared understanding on NFRs to rework tasks. 
In an iterative, collaborative process with our partner organizations, we performed an in-depth analysis of such tasks to identify reasons for lack of shared understanding, including whether it was \textbf{accidental} or \textbf{essential}, and which practices to employ to avoid lack of shared understanding.

\subsection{Collaborating Organizations}
We identified three small, agile organizations using CSE and cloud-based platforms through local contacts and trade shows. 
Our organizations shared development artifacts, granted access to employees, and participated in focus groups.
For this paper our organizations are referred to as Alpha, Beta, and Gamma.
Each organization develops in-house software utilizing cloud-based platforms, which contributes to their primary source of revenue.
Each organization has been in business for 8-10 years and has 40-100 employees.
Each organization implements CSE, including automated builds and testing and varying levels of automated deployment.
One important trait is that the bulk of their respective development teams are co-located, with few exceptions.
This distinction is important, as it eliminates the plethora of problems associated with globally distributed software development, in particular RE \cite{damian2006guest, herbsleb2003empirical, paasivaara-gsd}.

Alpha works in the crowdsourcing industry collecting large amounts of data on a daily basis.
Beta provides an e-commerce platform for customers distributed worldwide.
Gamma is an online content provider, including advertisements.

\subsection{Preliminary Study}
We first sought to understand which NFRs were relevant to our organizations and what processes and tools they employ to handle NFRs. 
One author spent multiple days over a few months embedded at each organization, learning about their products, processes, and business \cite{potts_research}. 
Through our immersive visits we spoke with 37 different team members from various departments.
Across our organizations, we spoke with 17 developers, 8 development managers, 5 product managers, 4 executives, 2 customer success specialists, 1 quality assurance member, and 1 director of sales.
As part of these informal conversations, we discussed shared understanding, NFRs, and rework, including how much each organization estimates they spend on rework, how they manage tacit knowledge, and NFRs that cause rework.
Our early immersive meetings and observations at our organizations informed our focused investigation into the lack of shared understanding of NFRs and its relationship to rework.

\subsection{Data Collection}To analyze lack of shared understanding of NFRs as it materialized in rework, we \hll{performed an in-depth} analysis of development tasks at each organization, \hll{where each task represented either a bug, feature, a story, or an epic, depending on the context at each organization.}
We analyzed and systematically coded these development tasks from task-management tools in relation to three elements: 1) NFRs, 2) lack of shared understanding, and 3) rework.
This coding was developed and validated in an iterative and collaborative process with our organizations \hll{over two full months. We indicate below the full time taken for specific analysis activities.}
\begin{figure}
    \centering
            \includegraphics[width=0.5\textwidth]{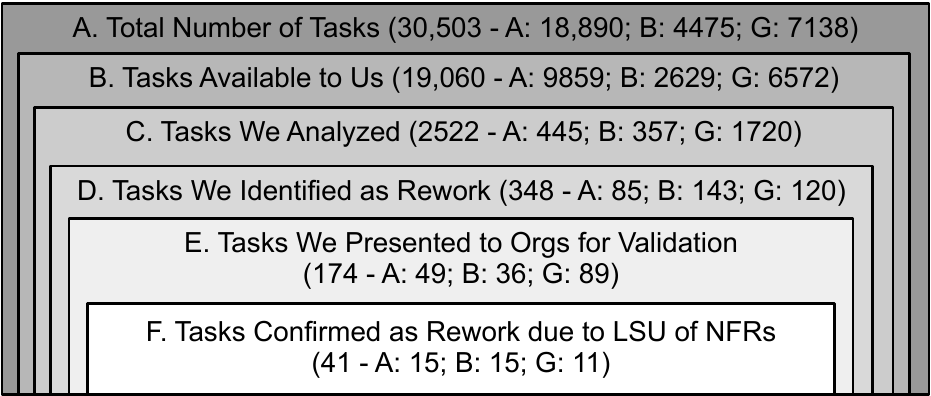}
    \caption{Sampling of rework tasks due to lack of shared understanding of NFRs (Total \# of tasks - \# of tasks at Alpha (A); Beta (B); and Gamma (G))}
    \label{fig:task-sampling}
\end{figure}

\subsubsection{Identification of NFR-related Development Tasks that Represented Rework}
Each organization granted us access to their respective task management tool (Jira) under a non-disclosure agreement.
We exported all available tasks, and associated attributes through a script \hll{over one week and spent another week understanding the data.}
The resulting data contained 30,503 tasks across our organizations (Figure~\ref{fig:task-sampling} Box~A).
Due to the large number of tasks and limited time with each organization, we performed task sampling \hll{over four weeks} to achieve a more practical and manageable number of tasks for our analysis, as summarized in Figure~\ref{fig:task-sampling}.

With the help of each organization, we removed tasks of projects with employees \emph{unavailable} for our focus groups, reducing the number of tasks to 19,060 (Figure~\ref{fig:task-sampling} Box~B). 
We then narrowed the analysis to the most recently closed tasks in the last 12 months, totalling 2522 tasks (Figure~\ref{fig:task-sampling} Box~C), to ensure that our validation of NFRs and rework with employees was possible. 
\hll{The vast majority of the four weeks was spent on identifying} which of these 2522 tasks represented rework, we 1) analyzed each attribute of a task (title, summary, description, comments, assignee, date, estimated time) and identified keywords such as 'refactor', 'rework', 'improve' and 2) examined references to previous tasks. 
This analysis resulted in 348 tasks that \emph{we believed} represented rework (Figure~\ref{fig:task-sampling} Box~D).

\subsubsection{1st Member Checking and Validation of Rework Tasks}
\label{sec:expl_stdy_rslts}
We then confirmed if each task actually represented rework and, more importantly, if that rework was due to lack of shared understanding of NFRs. 
We held a number of meetings at each organization to discuss our selection of tasks; these meetings were always attended by 1) two researchers, 2) one developer, and 3) one manager, where each employee had a considerable amount of breadth and depth of technical knowledge. 
We started off by clarifying the definition of NFRs, shared understanding, rework, and the relationship between the three to ensure consistency.
Due to partner time constraints, we discussed only 174 of the 348 tasks \emph{we} identified as rework (Figure~\ref{fig:task-sampling} Box~E).
Overall, our organizations confirmed 41 tasks as rework due to lack of shared understanding of NFRs (Figure~\ref{fig:task-sampling} Box~F).
These 41 tasks were used as the basis for our in-depth analysis through further focus group sessions at each organization. 

\subsubsection{Focus Group Sessions on Rework Associated with Lack of Shared Understanding of NFRs}
We scheduled focus group sessions at each organization to discuss each organization's share of the 41 rework tasks. 
We examined the factors contributing to lack of shared understanding (RQ1), which NFRs were more prone to lack of shared understanding (RQ2), and what amount of lack of shared understanding could be avoided (\textbf{accidental}) versus \textbf{essential} in the complex relationship with rework (RQ3). 
Each focus group lasted three hours and, similar to the validation of rework tasks, involved 1) two researchers, 2) one developer, and 3) one manager.
We performed 2 focus group sessions at each organization.
The base set of questions for each of the 41 tasks is in Table~\ref{tab:interview-questions}.
\hll{We clarified the definition of NFRs, shared understanding, rework, avoidability, and the relationship between the four in a similar fashion to our 1st member checking phase.
For example, we ensured that each participant equally understood avoidable in the sense that the organization could have prevented the lack of shared understanding by taking action prior to the rework task arising.}
These sessions were, with permission, audio-recorded. 
\begin{table}
    \centering
    \caption{Base Focus Group Questions}
    \label{tab:interview-questions}
    \begin{tabular}{rp{7.5cm}}
    \toprule
    \textbf{Q\#} & \textbf{Question} \\ 
    \midrule 
    Q1 & Do you know when/why/how the rework in this task was done? \\
    Q2 & Do you know when/why/how the original work leading up to this task was done? \\
    Q3 & Which NFR or NFRs do you associate with the lack of shared understanding of this task? \\
    Q4 & Do you know why there was a lack of shared understanding of the NFR for this task? \\
    Q5 & What could the organization have done to prevent the lack of shared understanding for this task? \\
    Q6 & Do you know why this action was not taken for this task? \\
    Q7 & Do you believe this lack of shared understanding was avoidable for this task? \\
    Q8 & How complex do you believe this task was? (Small, Medium, or Large) \\
    Q9 & How important was this task to the organization? (High, Medium, or Low) \\
    Q10 & Do you have anything else to add? \\
    \bottomrule
    \end{tabular}
\end{table}

\subsection{Data Analysis}
To answer our research questions, our data analysis triangulated contextual data from our preliminary study, quantitative data collected from each organization's task management tool, and qualitative data from the in-depth, focus-group sessions on the 41 rework tasks due to lack of shared understanding of NFRs. 
The key to our analysis of the rich, rework task data was the contextual, organizational knowledge we acquired during the preliminary study at each organization. 

To analyze the qualitative data, we transcribed the recordings of the focus-group sessions to obtain 41 transcripts\hll{, taking one week}. 
We performed thematic analysis \cite{cruzes_recommended_2011} \hll{spanning 4 weeks} on these transcripts. 
We used an open coding \cite{corbin1990grounded} approach to develop a purely inductive codebook \cite{cruzes_recommended_2011}, which minimizes a coder's ability to force a bias of any particular hypothesis.
In the initial coding phase, a transcript was independently coded by two coders, after which an agreement session was held to discuss the codes, consolidate the codebook, and to calculate Cohen's kappa coefficient. 
We continued until the kappa value stabilized above 0.6 (substantial inter-rater reliability agreement \cite{landis1977measurement}), which occurred at the 20th transcript. 
Each of the remaining 21 transcripts was coded by a single coder and reviewed by a different coder.
Throughout all coding we used the constant comparison method; codes were added, removed, and merged based on the discussions between the coders. 
Our final codebook \hll{took two weeks to compile and }had 48 codes (85\% of the codes were used in the first 10 transcripts). 
The complete codebook is in our replication package. 
To answer RQ1 and RQ3, we employed thematic synthesis \cite{cruzes_recommended_2011} \hll{over a two week period} to abstract themes from our derived codes. 
We constructed themes by contemplating how different codes may be combined to form overarching themes. 
Three themes emerged as factors contributing to the lack of shared understanding of NFRs (RQ1). 
Similarly, a number of themes emerged related to the \textbf{accidental} versus \textbf{essential} nature of shared understanding (RQ3), including practices to avoid a lack of shared understanding. 
We performed a 2nd round of member checking with our organizations to validate these themes.
For RQ2, we counted the number of references to particular NFRs for each task.

\section{Findings}
\label{sec:findings}

\emph{RQ1: What Contributes to Lack of Shared Understanding of NFRs?}
\label{sec:rq1}
We asked participants if they knew why there was lack of shared understanding of NFRs for each of the 41 tasks (Q4).
In our data analysis three themes emerged: fast pace of change, lack of domain knowledge, and inadequate communication.
The themes and associated codes are in Table~\ref{tab:reasons-themes}.

\begin{table}
    \centering
    \caption{Summary of Factors Contributing to LSU of an NFR (RQ1)}
    \label{tab:reasons-themes}
    \begin{tabular}{rl}
    \toprule
    \textbf{Theme}  & \textbf{Associated Codes} \\
    \midrule
    \multirow{6}{*}{\textbf{Fast Pace of Change}} & EducationalRework \\
    & JustGetItToWork \\
    & LackOfResources \\
    & LackOfTests \\
    & ThirdPartyIntegration \\
    & TradeOff \\
    \midrule
    \multirow{5}{*}{\textbf{Lack of Domain Knowledge}} & BusinessContext \\
    & DifferentPriorityScale \\
    & ChangeHereBreaksThere \\
    & ChangingRequirements \\
    & LackOfKnowledge \\
    \midrule
    \multirow{3}{*}{\textbf{Inadequate Communication}} & Ambiguous   \\
    & Communication \\
    & InformationOverload \\
    \bottomrule
    \end{tabular}
\end{table}

\subsubsection{Fast Pace of Change (33 of 41 tasks)}
When an organization is moving rapidly there is little emphasis on NFRs due to the immense pressure to release a product, e.g. a manager at Beta discussing why \nfr{usability} is not on the forefront \q{time constraints, had to be out. When it was originally written we just needed it to work.}
Similarly, when considering \nfr{deployability}, a manager at Alpha said \q{when you're first starting out it's move fast and break things; get things out the door. There's fallout, could be minimal fallout, but something you take a chance on.}
Both Beta and Gamma \hll{suffered from \nfr{extensibility} issues,} recently performed a rewrite, and optimism surrounds the ability to increase shared understanding of the new system, e.g. \qp{our new [redacted] so a lot of our debugging like live to on the server is going to go away so that will change if you'd come and interview me in two weeks and see a completely different conversation}{Developer at Beta}.

Lack of testing was also a result of moving too fast, as our organizations admitted to not performing adequate testing, of either FRs or NFRs due to insufficient resources (time \emph{or} people), e.g. \qp{we weren't doing any kind of testing, a/b testing or anything like that to assess \nfr{performance} or whether these changes were having a positive or negative impact}{Developer at Gamma}.

\subsubsection{Lack of Domain Knowledge (29 of 41 tasks)}
Domain knowledge is the business-specific context required to compete in a particular domain. 
Glinz and Fricker's paper on shared understanding \cite{glinz_shared_2015} mentions domain knowledge as a problem to building shared understanding.
Our participants were clear that understanding the market you are developing for is key to continued success for the entire organization, e.g., \qp{it's important to have that business context of how people are going to use it and that's probably the biggest stride being made, is just trying to expand internally what we know about business context so that new people coming in [are aware]}{Developer at Alpha}.
Lack of domain knowledge can cause lack of shared understanding of NFRs if an organization is entering a new, unfamiliar market where even a basic understanding would be beneficial, e.g. \qp{if we had known more about the industry, which I mean again even just a user of the industry}{Developer at Beta} with respect to having a narrow view and not consider \nfr{extensibility}.
Alternatively, lack of shared understanding could occur due to the unknown unknowns \cite{sutcliffe-unknowns}, e.g. \qp{lack of scope, we never considered that this would be a thing}{Manager at Beta} concerning the \nfr{scalability} of how one of their customer's would utilize a particular feature.

We also observed an organization's difficulty in adapting to new horizontal markets with respect to localization techniques that created \nfr{maintainability} issues, e.g. \qp{start an app early on for North American companies requiring one [format]. As your app becomes more popular you start getting requests for new countries that use [other] formats}{Manager at Beta}.
Even if domain knowledge is well known, the level of comprehension or understanding may change, which may bring new knowledge about a particular NFR, e.g. \qp{they changed because of increased comprehension of the problem space we were trying to solve}{Developer at Gamma} in regards to acknowledging they did not grasp a \nfr{extensibility} requirement.

We also found the priority of an NFR is a result of the level of shared understanding of that NFR.
Reaching consensus on the priority of an NFR may be difficult due to a discrepancy in perceived importance between different units within an organization, as a developer at Gamma stated \q{definitely NFRs tend to be lower priority, or at least perceived as by the business as low priority}.
Additionally, we observed assessing an NFR as low importance \emph{until} the NFR becomes such a problem that the system no longer functions, e.g. \qp{I would say \nfr{performance} is never a high priority until somebody says it doesn't work for me in which case you've graduated from an NFR to it is not functioning for me}{Developer at Beta}.

Finally, organizations also face difficulty when dealing with complicated NFRs, such as \nfr{privacy} \cite{li2019continuous}.
Our organizations are particularly concerned about \nfr{privacy} as they are collecting and managing a lot of customer data. 
\nfr{Privacy} can be challenging as a developer may not have legal expertise to determine the correct course of action to comply with privacy law. 
Hence, legal consultants are frequently relied upon to provide guidance. 
However, despite legal consultation, Gamma's compliance with GDPR was hindered due to the inexperience of the developers.
Even if a developer acquired knowledge of the GDPR, there was no systematic method to develop shared understanding.

\subsubsection{Inadequate Communication (15 of 41 tasks)}
\hll{Developers told us that,} due to the implicit and cross-cutting nature of NFRs, they were not aware of the value of communicating NFRs until it was too late, e.g. a manager at Alpha explained \q{nobody explained to him what \nfr{reliability} was} and \q{then somebody gave them a code review or whatever and said, oh you should have done it this way.}
\hll{Developers explicitly reported communication problems.
For example, developers} working on the same problem, almost simultaneously, in isolation of each other but not communicating the \emph{right} information, which created \nfr{maintainability} issues e.g. \qp{two people [developing] independently, they talk to each other well, but you're still going to approach problems differently and hit different areas of the system and not know everything that's going on. So there was no notion of or expectation of \nfr{maintainability}}{Manager at Beta}.

Communication is the key to ensuring all developers have a shared understanding of NFRs; however, \hll{communication is often lacking}, e.g. \qp{it's just a \nfr{performance} matter, but yeah, just the people involved were unfamiliar with what they're using}{Developer at Beta}.
Communication is also time sensitive, it might \qp{take a little while before they sort of enter the common knowledge of the company}{Developer at Alpha} according to a developer at Alpha in reference to \nfr{extensibility} and \nfr{maintainability}.

In other instances, a developer makes a false assumption that everyone else has the \emph{same} understanding of terminology, knowledge, or perception of an NFR and the need to communicate the NFR is overlooked.
For example, the definition of quality and releasable is different between \emph{two} developers, \qp{it's a lack of understanding of like what is quality software and what does releasable mean?}{Manager at Gamma}.

Communication break-downs can also lead to challenges in disseminating information.
For example, \nfr{extensibility} and \nfr{performance} suffered at one organization due to \qp{two different developers building the front-end and the back-end at different times. Only like a week or two apart, but they weren't communicating well enough to each other and it wasn't enough like top-down planning of that}{Manager at Gamma}.
As a result, high priority rework occurred to rectify the effects of the lack of shared understanding.

\emph{RQ2: Which NFRs are Most Associated with Lack of Shared Understanding?}
We asked each organization what NFRs were associated with each task (Q3). 
The results of NFRs with at least six or more associated NFR-tasks are in Figure~\ref{fig:nfr_codebook}.
Each task could be associated with multiple NFRs.
The top NFRs: \nfr{reliability/availability}, \nfr{maintainability}, \nfr{usability}, and \nfr{extensibility}, respectively occurred in 17, 16, 15, and 14 tasks.
\pgfplotstableread[col sep = space]{data/nfr-count.csv}\nfrCount
\begin{figure}
    \centering
    \begin{tikzpicture}
    \definecolor{c1r1}{HTML}{6D463C}
    \definecolor{c1r2}{HTML}{6F8F95}
    \definecolor{c1r3}{HTML}{F4EB90}
    \begin{axis}[
        ybar stacked,
        xtick=data,
        bar width = 2.75mm,
        legend style={
            cells={anchor=west},
            legend pos=north east,
        },
        xtick pos=left,
        ytick pos=left,
        ymin=0,
        width=7cm,
        height=5cm,
        ymax=18,
        xmin=0.5,
        ylabel={Number of NFR-related Tasks},
        ylabel style={font=\footnotesize},
                xticklabels from table={\nfrCount}{NFR},
        ytick={0,2,4,6,8,10,12,14,16,18},
        legend style={font=\footnotesize},
        x tick label style={rotate=45, anchor=north east, font=\footnotesize, inner sep=0mm},
        y tick label style={font=\footnotesize},
        ]
    \addplot[ybar, fill=c1r1,draw=black!70] table [meta=NFR, y=Alpha, x=Number] {\nfrCount};
    \addlegendentry{Alpha}
    \addplot[ybar, fill=c1r2,draw=gray!80] table [meta=NFR, y=Beta, x=Number] {\nfrCount};
        \addlegendentry{Beta}
    \addplot[ybar, fill=c1r3,draw=gray!60] table [meta=NFR, y=Gamma, x=Number] {\nfrCount};
        \addlegendentry{Gamma}
    \end{axis}
    \end{tikzpicture}
    \caption{NFRs with $\geq$ 6 NFR-related tasks (RQ2)}
    \label{fig:nfr_codebook}
\end{figure}
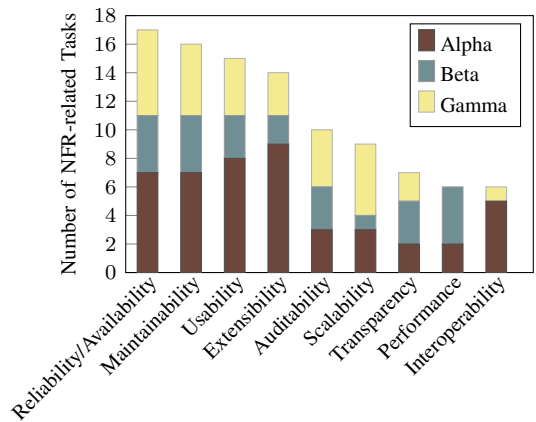

\emph{RQ3: What Amount of a Lack of Shared Understanding is Accidental versus Essential?}
Recall we asked participants whether the lack of shared understanding was avoidable (Q7).
We also asked participants how the lack of shared understanding could be avoided (Q5) and why this action was not taken (Q6).

Overall, the responses show 78\% of the lack of shared understanding of NFRs was considered \textbf{accidental} across our organizations.
The remaining 22\% represents a \textbf{essential} lack of shared understanding.

\emph{Practices to Avoid Lack of Shared Understanding} 
\label{sec:practices}
In response to Q5 and Q6, two primary practices emerged to avoid a lack of shared understanding: standards, and communication and documentation.
However, in many cases the focus groups with our research team was the first time the organization had realized the true cost of lack of shared understanding.

\paragraph{Shared Development Standards (22 of 41 tasks)}
Our organizations believed further adopting more development standards could build shared understanding, particularly for \nfr{maintainability}. 
This would prevent rework, e.g. \qp{any time you needed a place to put something we used to toss them on that big stack of unorganized classes until it became a giant object}{Developer at Beta}
The lack of shared understanding means the giant object must be refactored and manifested \nfr{maintainability} issues ultimately created functional problems, as noted by a developer at Beta, \q{the rework on it was not only to refactor it [...] but essentially to handle the type of structure used and [redacted] caused problems throughout the app.}

Standards can be applied to multiple facets of an organization, not limited to coding standards.
Our organizations are actively standardizing their development methodology. 
Gamma mandates a developer must create a task to record work exceeding one hour, to help foster a shared understanding. 
This can help a new employee assigned to improve the \nfr{deployability} of the infrastructure, who may not have the context of the original infrastructure. 
If the original developer is unavailable, the assigned employee has no shared understanding (e.g. why was Jenkins modified?).
Similarly, a manager at Beta maintains that creating a standardized development process can \q{create more shared understanding in more advanced design up-front} to help avoid \nfr{usability} issues.

\paragraph{Adequate Communication and Documentation (19 of 41 tasks)}
Unsurprisingly, inadequate communication was one of the main contributors to lack of shared understanding, an obvious solution is to have an adequate amount of communication; however, more surprisingly is that this was not occurring often enough.
We observed communication issues between developers and a) other developers, b) support analysts, c) testers, d) product managers, and e) managers.
For example, Gamma experienced a situation where \q{[developer] was off on his own and he wasn't interacting with the development team or the project management team [and was short-sighted]}, causing major \nfr{maintainability} challenges. 
A manager at Gamma said the \nfr{maintainability} issue could be resolved by \q{better coordination and communication}, as the developer was a contractor and isolated from the rest of the development team and \q{there was no documentation about it.}
Additional documentation, code review, or walk-through was all that was needed to avoid the lack of shared understanding.

Communication across organizational roles is essential to reduce lack of shared understanding of NFRs.
For example, Alpha had a lack of shared understanding of \nfr{deployability} due to miscommunication between management and developers, \q{[we are] handing them off to a specific account manager and teaching them how to work through the system.}
Additionally, a manager at Gamma indicates \q{more collaboration to begin with} may be necessary to alleviate issues associated with \nfr{maintainability}, \nfr{extensibility}, and \nfr{scalability}.
In Gamma's case, the issue arose because there was a lack of communication between a developer working on a feature and the rest of the development team.
In consequence, the developer was unaware of the expectations in conjunction with rest of the software. 
This issue could be mitigated if more upfront communication was established and the developer had a shared understanding of the development team's practices. 
Documentation is communication that takes the form of explicit shared understanding.
For example, a developer from Alpha believes \q{documentation [is desired] because [it gives] you understanding what's going on during that time frame}.
Without documentation, tracking decisions and actions may be difficult. 
Moreover, on-boarding and training is challenging without documentation serving as a guide.

\section{Discussion}
\label{sec:discussion}

In this work we aimed to empirically investigate shared understanding in RE. 
In particular, we studied the lack of shared understanding of NFRs and its effects as traced in the form of rework in software projects. 
Using mixed-methods, we studied three small, agile organizations practicing CSE and using cloud-based platforms. 
We found some NFRs (RQ2) (\nfr{reliability/availability}, \nfr{maintainability}, \nfr{usability}, and \nfr{extensibility}) are more prone to lack of shared understanding in these organizations, and the majority of this lack of shared understanding (78\%) was \textbf{accidental}, i.e. avoidable, whereas 22\% represented \textbf{essential}, i.e. not avoidable, lack of shared understanding (RQ3). 

From our analysis three factors contributing to the lack of shared understanding (RQ1) emerged: fast pace of change, lack of domain knowledge, and inadequate communication. 
We start our discussion of these findings in the context of our organizations' operating environment. 
In particular, these were agile organizations practicing CSE, which \hll{we believe may pose additional} challenges to the treatment of NFRs. 
We aim to shed some light on the complex relationship between shared understanding, rework, and CSE. 

We also identified two practices (RQ3) (shared development standards and adequate communication and documentation) whereby organizations may benefit in building a shared understanding and awareness of NFRs. 
However, organizations have a fixed set of resources and thus need to strategically prioritize when to build a shared understanding against developing new features. 
Our next point in discussion will be on \emph{when} and \emph{how} organizations could employ these strategies to build shared understanding in the context of CSE. 

\hll{We believe} these findings have important research implications for the intersection of shared understanding of NFRs and CSE. 
In addition, our final discussion point refers to practical implications from our findings, including adaptations \hll{we believe may} enable a more proactive approach to avoid \hll{rework due to an} \textbf{accidental} lack of shared understanding.

\subsection{Shared Understanding and CSE}
At first sight, CSE advocates for ongoing customer feedback and rapid iteration cycles \cite{fitzgerald2017continuous} that should result in a heightened understanding of customer requirements. 
In our deeper empirical investigation, our study reveals \hll{the potential for} a much more complex relationship between lack of shared understanding (as materialized in rework) and practices of CSE. 
Our findings suggest that \emph{CSE \hll{may hurt the} shared understanding of NFRs and lead to rework}. 

One expected benefit of CSE is an improved validation of requirements, due to rapid customer feedback \cite{kersten19}. 
A quick feedback loop is associated with lowering the impact of lack of shared understanding \cite{glinz_shared_2015}. 
However, as NFRs are cross cutting and difficult to decompose into sub-components that can be completed within a short iteration \cite{bellomo_evolutionary_2014}, an NFR may take days, weeks, or even months to evaluate.
For example, at Alpha there was rework due to a lack of shared understanding of \nfr{privacy} that was untestable, \q{we never really know what we're testing because it requires a host of real users hitting it to get an accurate picture of your hits over the last couple of weeks, developers won’t have weeks of data.}
Beta had a similar experience where \q{a new FR came in to integrate a third-party service, which we dutifully did, but it cratered system \nfr{performance}. Embarrassingly, there were so many other changes going on at the same time (some our own and some seasonal) that it took us a few weeks to discover and then track down the problem.}
For our organizations, \hll{we hypothesize that} CSE hurts the shared understanding of NFRs, much like how agile RE practices increase the risk of overemphasis of FRs \cite{ramesh_agile_2010}.
A notable complication is CSE has a broader scope than agile, as not only do you need to slice NFRs \cite{ernst_enabling_nodate}, but you need to subsequently ship them in rapid, frequent deployments \cite{savor_continuous_2016}.

We found NFRs were de-prioritized in CSE, in part due to the frequent release of FRs being a top priority and a lack of shared understanding, around either domain or technical knowledge.
NFRs may not get prioritized until the right (profitable) customer complains, e.g. \qp{flagship user of this product, deemed very important}{Manager at Beta}.
At our organizations, product managers are often tasked with creating, prioritizing, and validating tasks.
Tasks were typically only created for FRs, and the NFRs are usually implicit, and in some cases as what Glinz and Fricker call `dark information', i.e., relevant but unnoticed knowledge \cite{glinz_shared_2015}.

Another claimed benefit of CSE is the removal of intra-organizational barriers, such as between product managers, DevOps, and developers \cite{fitzgerald2017continuous}.
These barriers are prohibitive to both adequate communication and acquiring domain knowledge.
By tearing down these barriers, CSE should help alleviate the difficulty associated with cross-cutting NFRs that require input from multiple individuals and roles across an organization.
However, we found that a lack of shared understanding existed across organizational departments and roles and NFRs remain difficult for non-developers to grasp. 
Product managers in particular ceded control of NFRs to developers, relying on intuition to verify the NFR was achieved: \qp{we're going to see this thing before it goes out the door and we'll know something feels wrong in terms of a NFR}{Manager at Beta}.
In particular, \nfr{performance} and \nfr{scalability} cannot be validated by a single product manager ``knowing when something feels wrong,'' when they actually require thousands of customers to verify.

Our findings also suggest a disconnect between development and product management at these organizations practicing CSE, where a product owner would typically bridge the knowledge gap.
Unlike agile development, CSE does not define a product owner role, who would have substantial technical \emph{and} customer domain knowledge. 
At our organizations the product manager role was responsible for writing the tasks and deciding each task's priority.
However, due to this disconnect NFRs requiring substantial technical \emph{and} customer domain knowledge, such as \nfr{security} and \nfr{privacy} \cite{compagna_how_2009}, were often neglected or underspecified \cite{femmer_rapid_2017} leading to a lack of shared understanding.
We found the primary difference between an agile product owner and our organization's product managers is the insufficient level of technical knowledge; therefore, CSE \hll{may be} left without the benefit of a true product owner.

On the positive side, CSE relies heavily on writing and maintaining numerous test cases (a form of shared understanding) that are part of the automated CSE build \cite{glinz_shared_2015}. 
However, we found that our organizations did not heavily invest in substantial testing efforts, even for FRs.
Furthermore, NFRs are difficult to test \cite{borg_bad_2003, jiang_co-evolution_2015} and may require specific architectural considerations to be automatically tested \cite{ameller2012non}, therefore NFRs are largely left as a manual task in CSE \cite{caracciolo_how_2014}.
The lack of NFR tests could be due to a lack of shared understanding (i.e. they don't know which ones to test); alternatively, it could contribute to lack of shared understanding of NFRs, (i.e. tests themselves represent shared understanding anyone can read, modify, or execute).
\hll{For example, \nfr{interoperability} may be degraded due to inadequate tests \qp{the root of it is we built one half (front-end) without building the other (back-end)}{Developer at Beta}.}
Even if an NFR is not fully tested as part of the CSE pipeline, the organizations are not realizing the full benefits of the CSE pipeline due to a lack of tests.
One key to building a shared understanding of NFRs through tests in CSE is to, at the very least, have imperfect tests \cite{fowler-continuous-int}; these imperfect tests are the first step in acknowledging and bringing awareness to the value of the NFR and start fostering a shared understanding.

\begin{mybox}[Research Implication 1]
\hll{Our evidence suggests that one possible side-effect of CSE is a decrease in the shared of understanding of NFRs that leads to rework.
There is a need for further exploratory studies to confirm this, as well as defining roles, methods, and tools to identify and mitigate the potential lack of shared understanding in CSE. 
How could revised methodologies leverage a partnership and mutual respect between product managers and developers to ensure adequate dissemination of NFR and domain knowledge in CSE organizations? }

\end{mybox}

\subsection{What Triggered Our Organizations to Build a Shared Understanding?}
\label{sec:disc_when}
Our organizations were not actively trying to build a shared understanding.
In the CSE approach, they try something and assess the viability once it is deployed in the CSE pipeline. 
If it was not viable (e.g., did not increase revenue) they simply pivot to other features and tasks.
As we discussed in the preceding section, our organizations acknowledge that rework due to an \textbf{accidental} lack of shared understanding of NFRs is a problem. 
However, the question of \emph{when} to build a shared understanding of NFRs is tricky. 
Rather than proactively doing this, our organizations relied on reactive stimuli.
They used different triggers to identify the need for building shared understanding of NFRs. 
These triggers were regulatory requirements, accumulating technical debt, needs of important customers, and disruption of service.

\subsubsection{Regulatory Requirements}
A core part of each organization's business was to collect and use customer data.
The incentive for building a shared understanding of regulatory NFRs, such as \nfr{privacy} and \nfr{security}, is to reduce the potential liability.
These laws or regulations, such as the GDPR, are important and comprehensive, as non-compliance could result in crippling financial or legal penalties \cite{li2019continuous}.
Explicit regulatory policies, such as \nfr{privacy}, can be extracted, visualized, and re-published to ensure a higher quality of shared understanding between various stakeholders \cite{silva_improving_2020}. 

For Gamma, the GDPR represented making substantial adjustments not only to the organization's business, but also raising the level of shared understanding of \nfr{privacy}.
Instead of \nfr{privacy} being an important, but possibly unconsidered quality when faced with time constraints, Gamma needs to ensure that everyone (i.e. not only developers) in the organization has a shared understanding of the GDPR. 
For instance, third-party services and libraries must be verified for GDPR compliance \cite{li2019continuous}.
Employees need to understand the risk of finding and using an external library without first vetting the library for \nfr{privacy} considerations.

\subsubsection{Accumulating Technical Debt}
\hll{At our organizations,} another trigger was to build a shared understanding of NFRs when the organization had incurred significant technical debt as a result of \textbf{accidental} lack of shared understanding.
This often occurs when technical debt is ignored, \hll{potentially} exacerbated by \hll{the fast pace of} CSE, and the belated response by technical staff to recognize the need to change \cite{ernst2015measure}.
Refactoring software has many benefits, in particular for NFRs, such as increased \nfr{reliability}, \nfr{maintainability}, and \nfr{reusability} \cite{kolb2006refactoring, kim_refactoring}.
However, \hll{we believe the} benefit is usually technical in nature (e.g. the system becomes more \nfr{reliable}) and the impact of refactoring on shared understanding is yet unknown.
Due to the amount and constraints of technical debt, two organizations (Alpha and Beta) rearchitected their entire systems from scratch.
The effect of this rearchitecture was two-pronged.
First, each organization had the opportunity to build an entirely \emph{new} architecture.
The new architecture encouraged developers to openly discuss and evaluate NFRs, such as \nfr{availability}, \nfr{scalability}, \nfr{deployability}, \nfr{scalability}, and \nfr{maintainability}, and developed a well-understood, shared understanding of these NFRs.
Second, each organization was able to build a shared understanding of not-well-understood, legacy NFRs from the \emph{previous} system in achieving feature parity between the two. Thus, we found that refactoring can \hll{potentially} \emph{increase} the shared understanding of NFRs.

\subsubsection{Needs of Important Customer}
As our organizations have a broad set of customers, each of which must be satisfied, it is vitally important to identify and prioritize perspective customers \cite{saiedian2000requirements}; however, often the focus is on FRs.
At two of our organizations (Alpha and Gamma) we found that a shared understanding was triggered as a result of some input from an important customer or client, usually in the form of a complaint.
At Alpha an important customer was using a very large dataset, which Alpha had not considered or tested.
This large dataset caused \nfr{performance} issues due to a lack of \nfr{scalability} in the architecture and resulted in a high priority rework task.
As a result of the high priority, a shared understanding was developed for both \nfr{performance} and \nfr{scalability} across a large number of developers.
The shared understanding made developers aware that the original implementation lacked \nfr{performance} measures (and how to fix them) and brought attention to the various scales with which customers were using their system.
Had this customer not been as important Alpha may not have focused the necessary resources to build this shared understanding.

\subsubsection{Disruption of Service}
Disruptions of service are never a desired trait, as a recent disruption at Amazon was estimated to cost \$66,240 per minute of downtime \cite{amazon_downtime}.
Disruption of service usually involves a loss of functionality, and subsequent rework to patch the problem. 
The silver lining in a disruption is that we found our organizations focused on the lack of shared understanding of NFRs when a disruption of their service occurred.
Service disruptions are usually tightly related to an NFR, \nfr{availability} or \nfr{scalability} issues. If the effect of the disruption was small (i.e. one small customer or one small rarely used component) then the shared understanding was not widespread.
\hll{Participants in our study claimed that} incident post-mortems as an excellent example of how service outages can (somewhat painfully) force organizations to confront gaps in shared understanding. 
For example, Alpha uses post-mortems to increase understanding to focus on the most important aspects of the service. 

\begin{mybox}[Research Implication 2]
\hll{The fast paced CSE environment is one possible explanation why an organization would be more reactive to triggers, such as technical debt, as opposed to proactive in building a shared understanding of NFRs.}
We believe there is a need for further empirical evidence and examples of triggers, both reactive and proactive, and methods and tools to help identify triggers as opportunities to build shared understanding.
How effective are triggers in building shared understanding? How can we identify reactive triggers before rework occurs and use them as proactive triggers? Is there a difference in the level of shared understanding between a reactive and proactive trigger?

\end{mybox}

\subsection{Implications to Practice: How Could Organizations Build a Shared Understanding in CSE?}
Upon encountering a trigger (i.e. regulatory requirements, rectifying technical debt, important customer complaints, and disruptions of service) for building a shared understanding, the next step is actually initializing a response.
However, our organizations are largely reactive to rework due to \textbf{accidental} lack of shared understanding (e.g., responding to a disruption or customer complaint), as opposed to being proactive in building a shared understanding.
RE research has shown that a certain amount of proactive RE is valuable \cite{damiantse2006}.
Furthermore, the participants in our study suggested two practices that could help them build a shared understanding: shared development standards and documentation and communication. 
We discuss practical implications \hll{as suggested by our empirical findings and which we believe would enable an} organization to achieve a more proactive RE approach to their software projects.

While shared standards for documenting NFRs across an organization are necessary to build a shared understanding \cite{de_boer_writing_2009}, we do not believe the documentation needs to be exhaustive or complete.
Much like agile encourages ``working software over comprehensive documentation'', there is a minimum level of NFR documentation required to achieve a shared understanding in CSE.
An organization must strike a balance to meet this minimum, and sufficient, level of NFR documentation \cite{clear2003documentation}.
While determining what is considered a sufficient level of documentation might be different at each organization \cite{santos2015fostering}, a minimum level could be as simple as writing the NFR as part of the acceptance criteria in a task to a more involved characterization \cite{chen_characterizing_2013}.

An organization should first select a single, vitally important NFR, such as \nfr{performance} or \nfr{reliability} based on their own context.
In the most simplistic manner possible, the NFR must be documented to ensure cross-functional visibility.
All of our organizations maintain some form of dashboard that helps increase cross-functional visibility, thus we believe these dashboards are able to act as a form of documentation and be beneficial in increasing the cross-functional visibility of the NFR.
However, documentation on its own is not sufficient for building a shared understanding. 

Communication is a well known and studied knowledge management practice \cite{desouza2006knowledge, ko2007information}, thus the next logical step is to share this minimal NFR documentation across roles and departments in a CSE practising organization, which may require cognitive understanding \cite{junqueira_barbosa_empirical_2014}.
Establishing a cognitive shared understanding involves identifying insufficiency in an organization's understanding and rectifying misunderstandings \cite{junqueira_barbosa_empirical_2014}. 
The method of communication to establish this cognitive understanding could materialize in a variety of forms, including face-to-face conversations, emails, or corporate branding.

Once a standard level of NFR documentation has been developed and disseminated across all roles and organizational departments, then the value and awareness of the NFR must be incentivized to achieve buy-in from all roles and organizational departments.
This may be achieved through executive sponsorship through a top-down approach.
In contrast, documentation can also be achieved from the grassroots level, where developers initialize and progress towards disseminating knowledge between roles in the interest of self-help. 

Finally, the NFR itself should be verified as part of the CSE pipeline.
A well-known solution \cite{glinz_risk-based_2008, bellomo_evolutionary_2014} is to ensure NFRs can be quantified, for example as response measures or acceptance criteria.
\hll{As part of their efforts for \nfr{auditability}, our collaborating organizations actively add metrics to catch errors and increase insight on their system.}
We recognize that some NFRs are more difficult to verify than others; although even \nfr{privacy} has been managed through custom tools that automatically tests for software deficiencies \cite{li2019continuous}. 
Once an NFR can be verified, the results should be publicly displayed throughout the organization, to continue and maintain a high level of awareness.
Furthermore, if a particular NFR is not satisfied an organization can adopt a similar practice from lean software development whereby all employees must halt production to fix an issue \cite{poppendieck2012lean}, which will help excite shared understanding across roles and departments.
This ensures that shared understanding is consistent and maintained across the organization, including new employees.
Ultimately, \hll{we believe that} building awareness and educating the organization regarding the value behind building a shared understanding of NFRs, including priority, is paramount to software success.

\section{Threats to Validity}
\label{sec:threats}
To ensure validity of our qualitative research, we provide a replication package (\replicationPackage{}), including statistics of the analyzed tasks, codebook, and inter-rater agreement sessions. 
Unfortunately, due to non-disclosure and ethics agreements we are unable to reveal full transcripts.

We discuss threats present in data gathering, data analysis, and the results. 
When gathering data, we mitigated the effects of respondent bias (the observer effect) through two steps.
First, we assured the participants that our presence was not to critique them, but rather to acquire insight about lack of shared understanding of NFRs in their respective organizations.
Second, we discussed with two participants simultaneously for each focus group and encouraged the participants to not only talk to us, but also to each other. 
\hll{To ensure the participants had a shared understanding of particular terms (e.g. NFR, shared understanding, avoidable) we discussed the definitions prior to our questions.}
\hll{For construct validity, while interviewing participants we defined and used the term avoidable lack of shared understanding; however, in our theoretical conceptualization we use the creative analogy of \textbf{accidental} or \textbf{essential}.}
We also used transcription tools to help transcribe the audio of the discussion. 
To ensure that the transcription was accurate, a human listened and verified each audio file. 
One limitation may be our task sampling, as we did not analyze every task.
This was due to the limited time we had with each organization and restricted access to the data. However, the organization participants felt our selected tasks were representative of tasks in their context.

With respect to credibility in data analysis, 
we conducted pair coding until we reached a point where we consistently achieved a moderate to substantial level of inter-rater agreement.
Second, we conducted two rounds of member checking with our organizations and participants. 
The first round of member checking involved the organizations confirming the 41 tasks were rework as a result of lack of shared understanding of an NFR.
The second round of member checking verified our findings and themes, including factors contributing to and practices to avoid lack of shared understanding of an NFR.
We elicited ordinal feedback (Strongly Disagree-Strongly Agree) on each of our themes and practices. For all three themes and three practices, the respondents had a median score of ``Agree''.
\hll{Another threat is the relatively small sample size (41 tasks), partly due to the considerable amount of work required for each task, and the limited time with our organizations.
Finally, for conclusion validity we recognize that our research implications are phrased as hypotheses, \emph{not} factual conclusions, requiring further research investigations.}

\section{Conclusion}
\label{sec:conclusion}
Effectively building and managing a shared understanding of requirements is key to successful software projects. NFRs pose additional challenges given their cross-cutting nature that makes them more difficult to handle, particularly in CSE deployments.

Evidence from our study suggests that CSE negatively affects the shared understanding of NFRs, resulting in important implications to research that should reconsider RE in the context of CSE. Our research renews and refocuses our awareness on the complex relationship between rework and shared understanding in CSE.

With respect to practical implications of our study, CSE does not absolve an organization's responsibility to maintain a shared understanding of NFRs, including tracking and evaluating how much rework occurs due to an \textbf{accidental} lack of shared understanding.
An organization may also proactively engage employees in building a shared understanding of NFRs. 
To a large extent, the effort that organizations expend on building shared understanding or mitigating the effects of a lack of it will depend on an organization's ability to weigh the cost of the \textbf{accidental} lack of shared understanding of NFRs relative to its need for functional delivery.  Development of methods for such assessment, and to address the other research implications mentioned above are worthy of future work.

\section*{Acknowledgments}
\label{sec:ack}
We thank our three partner organizations and employees for their time and collaboration. 
We acknowledge Omar Elazhary, Austin Lee, and Derek Lowlind (University of Victoria) for their assistance in our study.
Our research was supported by grant NSERC-CRD 535876.

\bibliographystyle{IEEEtran}
\bibliography{main}

\end{document}